\documentclass[acmsmall,nonacm]{acmart}

\usepackage{listings}
\usepackage{xcolor}
\usepackage{enumitem}
\usepackage{amsmath}
\usepackage{wrapfig}
\usepackage{caption}
\definecolor{codeblue}{rgb}{0,0, 1}
\definecolor{codegreen}{rgb}{0,0.6,0}
\AtBeginDocument{%
  }

\setcopyright{acmlicensed}
\copyrightyear{2018}
\acmYear{2018}
\acmDOI{XXXXXXX.XXXXXXX}

\acmJournal{JDS}
\acmVolume{37}
\acmNumber{4}
\acmArticle{111}
\acmMonth{8}

\begin{document}

\title{Increasing the Expressiveness of a Gradual Verifier}

\author{Priyam Gupta}
\affiliation{%
  \institution{Purdue University}
  \country{United States}}
\email{gupta751@purdue.edu}

\begin{abstract}
  Static verification provides strong correctness guarantees for code; however, fully specifying programs for static verification is a complex, burdensome process for users. Gradual verification was introduced to make this process easier by supporting the verification of partially specified programs. The only currently working gradual verifier, Gradual C0, successfully verifies heap manipulating programs, but lacks expressiveness in its specification language. This paper describes the design and implementation of an extension to Gradual C0 that supports unfolding expressions, which allow more intuitive specifications of recursive heap data structures. 
\end{abstract}

\acmArticleType{Review}

\maketitle

 
\section{Introduction}
\textit{Gradual verification} \cite{Bader18,wise2020gradual} is a technique that combines compile-time and run-time verification techniques to allow incremental specification and verification of programs. Writing full specifications for a static verification tool is an all-or-nothing complicated process; gradual verification removes the burden of writing specifications for a complete system by allowing users to verify programs with missing/incomplete specifications. 
It allows formulas to be either precise (complete) or imprecise. Non-contradictory strengthening of imprecise formulas marked with ? (e.g. \texttt{? \&\& x.f == 2}) can be assumed statically, along with run-time checks added to ensure soundness. 

Gradual C0 \cite{divincenzo2022gradual} is a practical and efficient gradual verifier, built on top of the Viper \cite{Muller16} static verifier. Gradual C0's back-end (called Gradual Viper) extends Viper's \textit{symbolic execution} based verifier to support imprecise formulas. Utilization of Viper's infrastructure simplifies the implementation of gradual verifiers for different front-end languages. Gradual C0's front-end chooses to target the C0 programming language \cite{arnold2010c0}, a safe subset of C designed for education. Gradual C0 alerts users (with static and dynamic errors) to true inconsistencies between their specifications and code, and suppresses static errors caused by missing specifications. 
Gradual C0 and Viper are based on \textit{implicit dynamic frames} (IDF) \cite{Smans09}, a variant of \textit{separation logic} \cite{Reynolds02}, used for verifying heap manipulating programs. They also support \textit{recursive abstract predicates} \cite{Parkinson2005SeparationLA,Smans09}, which along with IDF allows users to verify programs containing recursive heap data structures. A predicate is an abstraction for any assertion that may contain information like heap permissions and/or expressions. An accessibility predicate (from IDF), denoted as \texttt{acc(x.f)}, represents permission to access field \texttt{f} of object \texttt{x}.

One current limitation of Gradual C0 is that its specification language lacks support for Viper’s \texttt{unfolding p($\overline{e}$) in b} construct  \cite{Muller16} (henceforth called \texttt{unfolding} expression). This forces users to devise complex ways of specifying recursive heap data structures (see linked list example in Listings \ref{lst:label} and \ref{lst:label2}). 
A nontraditional implementation may also be necessary (e.g. storing subtrees' heights in each node of an AVL tree to specify balanced property).  
Note that \texttt{unfolding} expressions are side-effect free, and thus, also extensively used in pure functions to compute over recursive heap data structures (pure functions will be supported in a future extension to this work). 
An \texttt{unfolding} expression temporarily exposes a predicate body to \textit{frame} (prove ownership of) heap locations used in an expression. Predicate definitions are treated \textit{iso-recursively} \cite{Summers13}, so they need to be explicitly unfolded to access the assertion in their body.  Predicates may be recursively defined (like \texttt{sortedList} in \ref{lst:label} and \ref{lst:label2}) and so, are useful for verifying unbounded recursive heap data structures. 
\begin{figure}[!ht]
\begin{minipage}[t]{0.47\linewidth}
{\scriptsize\ttfamily
\begin{lstlisting} [caption=List sortedness currently specified., xleftmargin=2em, name=bst-bad, numbers=left, label={lst:label}]
struct Node { int data; struct Node *next; };
/*@
predicate sortedList(struct Node *this, (*@\colorbox{red!30}{int prev}@*)) =
  (this == NULL) ? ( true ) :
    (
      acc(this.data) && acc(this.next) && 
      sortedList(this.next, (*@\colorbox{red!30}{this.data}@*)) &&
      (*@\colorbox{red!30}{this.data >= prev}@*)
    ) ;
@*/
\end{lstlisting}
}
\end{minipage}\hfill
\begin{minipage}[t]{0.50\linewidth}
{\scriptsize\ttfamily
\begin{lstlisting}[caption=List sortedness specified with unfolding., xleftmargin=2.5em, name=bst-good, numbers=left, label={lst:label2}]
struct Node { int data; struct Node *next; };
/*@
predicate sortedList(struct Node *this) =
  (this == NULL) ? ( true ) :
    (
      acc(this.data) && acc(this.next) && 
      sortedList(this.next) && (this.next == NULL ||
       unfolding sortedList(this.next) in 
        (*@\colorbox{red!30}{this.data <= this.next.data}@*))
    ) ;
@*/
\end{lstlisting}
}
\end{minipage}
\label{fig:sortedList}
\end{figure}

\textbf{Example: specifying list sortedness.} A linked list is sorted if every node has a value less than or equal to its successor. Listing \ref{lst:label} shows that maintaining an additional auxiliary parameter \texttt{prev} (in a creative manner) is required to define a recursive abstract predicate that provides access to a sorted list (since permission for \texttt{this.next.data} cannot be made available without \texttt{unfolding}). On the other hand, in Listing \ref{lst:label2}, an \texttt{unfolding} expression is used to straightforwardly express relationship between consecutive nodes when defining the predicate.

\section{Approach}
Gradual Viper's algorithm has 4 major functions (modified from Viper) to evaluate expressions (\texttt{eval}), \texttt{produce} and \texttt{consume} formulas, and execute statements (\texttt{exec}). Producing a formula adds permissions and constraints while consuming a formula checks constraints and removes permissions from the symbolic state (used in symbolic execution based static verification). Our extension carefully considers each step of Viper's \texttt{eval} rule for \texttt{unfolding p($\overline{e}$) in b}, which performs the following in order: \texttt{consume} \texttt{p($\overline{e}$)}, \texttt{produce} \texttt{p($\overline{e}$)}'s body, \texttt{eval} \texttt{b} and lastly, reset the symbolic heap ($h$) to its version before the aforementioned \texttt{consume} call. 
$h$ tracks heap permissions and is a part of the symbolic state, which further tracks information such as path condition, variable mappings, etc. Gradual Viper additionally tracks run-time checks and has a separate optimistic heap ($h_?$) construct in its symbolic state. $h_?$ keeps track of optimistically assumed heap permissions; and thus, helps avoid duplicate run-time checks when those same heap locations are accessed later in the program. A naive design of \texttt{unfolding} expressions would reset both $h$ and $h_?$ (following Viper's design) to their respective versions before the \texttt{consume} call. However, a key realization of our design is that we can safely preserve some optimistic information (avoid a full reset for $h_?$) to minimize run-time overhead. We also extend Gradual Viper's branching strategies to work with \texttt{unfolding}.

\lstset{
  basicstyle=\scriptsize\ttfamily,
  keywordstyle=\color{codeblue},
  commentstyle=\itshape,
  showstringspaces=false,
  columns=flexible,
  breaklines=true,
  escapeinside={(*@}{@*)},
  morekeywords={predicate, int, NULL},
  xleftmargin=10pt
}
\setlength{\intextsep}{0pt}%
\begin{wrapfigure}{r}{0.50\textwidth}
\vspace{-0.05in}  
\begin{lstlisting} [caption=Method to insert a new node at the head of a sorted linked list, name=example, label = {lst:a_label} ]
struct Node* frontInsert(struct Node *head, struct Node *item) 
//@ requires ? && acc(item->next) && (head == NULL || 
     (*@\colorbox{green!30}{unfolding(sortedList(head)) in (head->data >=}@*) (*@\colorbox{green!30}{item->data))}@*);
//@ ensures sortedList(item); { 
  item->next = head;
  fold sortedList(item);
  return item; 
}
\end{lstlisting}
\end{wrapfigure}
Method \texttt{frontInsert} in Listing \ref{lst:a_label} is partially specified to show our \texttt{unfolding} evaluation strategy in a gradual context. The precondition (after \texttt{requires}) expresses the conditions required before calling \texttt{frontInsert} while the postcondition expresses the conditions satisfied after \texttt{frontInsert}'s execution. \texttt{?} in the precondition expresses imprecision and allows Gradual C0 to optimistically assume information. The \texttt{unfolding} expression (highlighted in green) helps express that \texttt{frontInsert} should only be called with arguments \texttt{head} and \texttt{item} such that \texttt{head->data >= item->data}.
Note, the \texttt{fold} operation in the method body consumes \texttt{sortedList(item)}'s body and produces \texttt{sortedList(item)}, which helps prove the postcondition. An \texttt{unfold} operation (not shown here) achieves the opposite. Note that \texttt{fold}/\texttt{unfold} statements are impure unlike \texttt{unfolding} expressions.

\textbf{Minimizing run-time overhead.} 
We choose to retain $h_?$ chunks during an \texttt{unfolding} expression evaluation differently based on whether the predicate body is precise or imprecise.


\begin{itemize}[leftmargin=12pt,topsep=2pt]
    \item \textit{Precise predicate body.} When producing the precondition, the state becomes imprecise. When evaluating the \texttt{unfolding} expression (in green), \texttt{sortedList(head}) is not in $h$ and is assumed optimistically, and temporarily unfolded, providing permission to \texttt{head->data}. Permission to \texttt{item->data} is optimistically assumed (added to $h_?$). Now, in our design, after the \texttt{unfolding} expression evaluation is complete, we choose to retain permission for \texttt{item->data} in $h_?$. This is safe because the predicate body is precise and so we concretely know that the optimistic permission (for \texttt{item->data}) was provided by imprecision (\texttt{?}) in program state outside of the \texttt{unfolding} expression evaluation. Now, when the \texttt{fold} operation checks for permission to \texttt{item->data}, its permission is found in $h_?$, and a run-time check (unnecessary) is not generated.
    \item \textit{Imprecise predicate body.} Consider replacing \texttt{acc(this->data)} with \texttt{?} in predicate \texttt{sortedList}'s body in Listing \ref{lst:label2}. Now, when evaluating the \texttt{unfolding} expression (in green), missing heap chunks (\texttt{head->data} and \texttt{item->data}) could be provided by \texttt{?} in either \texttt{sortedList(head)}'s body or in \texttt{frontInsert}'s precondition. In this case, our design conservatively assumes the former and removes newly assumed optimistic heap permissions after \texttt{unfolding} evaluation.
\end{itemize}

Additonally, in either case, we choose to add an instance of \texttt{sortedList(head)} to $h_?$ upon completing evaluation of the \texttt{unfolding} expression (in green). \texttt{sortedList(head)} is already framed by the \texttt{consume} call within the \texttt{eval} rule for \texttt{unfolding}, thus adding it to $h_?$ helps avoid any unnecessary framing checks later in the program (at the \texttt{fold} call in our running example). Our extension ensures that predicates added to $h_?$ are soundly tracked.

\textbf{Origin tracking.}
Gradual C0 tracks where branching in a program originates. This is useful because run-time checks need to be augmented with branch information so that they are only executed on specific execution paths. The variables used in a branch condition might change in value during symbolic execution. Origin tracking helps maintain the information needed to evaluate a branch condition at the correct point in the program at run-time.
When evaluating \texttt{unfolding} expressions, producing the predicate body might cause the program to branch if the body contains a conditional assertion (like in Listing \ref{lst:label2}). 
The origin for the branch in that case should be the respective \texttt{unfolding} expression. However, if the \texttt{unfolding} expression is part of a called method's signature or a folded/unfolded predicate's body, the origin should be where the method call or \texttt{fold}/\text{unfold}/\texttt{unfolding} of the aforementioned predicate resides in the program.

Viper's \texttt{eval} rule for \texttt{unfolding} expressions is modified for Gradual Viper to support the stated strategies for preserving optimistic permissions and origin tracking (see Appendix \ref{appendix}). 



\textbf{Future optimization.} With \texttt{unfolding} expressions support, it is possible to have multiple options for framing an expression. For example, consider the imprecise formula \texttt{? \&\& x.f == 2}. Either \texttt{p(x) \&\& unfolding p(x) in x.f == 2} (for some predicate \texttt{p(x)}) or
\texttt{acc(x.f) \&\& x.f == 2} could be the precise version (correctly framing \texttt{x.f}). Adding the correct chunk (to $h_?$) for framing will ensure efficiency since that heap chunk might be used again later. Currently, we simply assume the minimum option (\texttt{acc(x.f) \&\& x.f == 2}). We consider adding an `options heap' construct to track multiple framing options until one of them is concretely used in the program and added to $h_?$.
\section{Conclusion}
Supporting \texttt{unfolding} expressions allows Gradual C0 users to intuitively specify recursive heap data structures. Our design preserves viable optimistically assumed information to minimize run-time checks for efficient gradual verification.
Next steps include adding support for pure functions, creating new benchmarks that utilize these additional constructs, and formally proving soundness of this extension. A richer specification language will contribute to Gradual C0's goal of user-friendly verification and increased adoption of verification in developer workflows.
\newpage
\bibliographystyle{ACM-Reference-Format}
\bibliography{references}
\newpage
\appendix
\section{Appendix}
\label{appendix}
\begin{figure}[!ht]
\scriptsize\ttfamily
{
    \begin{align*}
    & \text{eval}(\sigma_1, \text{unfolding acc(pred}(\overline{e}), p) \text{ in } b, Q) = \\
    & \quad \text{if the unfolding is explicit then} \\
    & \quad \quad \text{eval}(\sigma_1, p :: \overline{e}, (\lambda \sigma_2, p' :: \overline{e'} \cdot \\
    & \quad \quad \quad bdy := \text{scale(pred}_{\text{body}}[\overline{x \mapsto e'}], p') \\
    & \quad \quad \quad \text{consume}(\sigma_2, \text{acc(pred}(\overline{e'}), p'), (\lambda \sigma_3, s \cdot \\
    & \quad \quad \quad \quad \colorbox{blue!20}{$\text{if }(\sigma_3.R.\text{origin} == \text{None}):$}\\
    & \quad \quad \quad \quad \quad \colorbox{blue!20}{$R' := \sigma_3.R\{\text{origin} := (\sigma_3, \text{unfolding acc(pred}(\overline{e}), p) \text{ in } b,  p' :: \overline{e'})\}$} \\
    & \quad \quad \quad \quad \colorbox{blue!20}{$\text{else:}$}\\
    & \quad \quad \quad \quad \quad \colorbox{blue!20}{$R' := \sigma_3.R$}\\
    & \quad \quad \quad \quad \text{produce}(\sigma_3\colorbox{blue!20}{$\{R := R'\}$}, bdy, s, (\lambda \sigma_4 \cdot \\
    & \quad \quad \quad \quad \quad \text{eval}(\sigma_4{\colorbox{blue!20}{$\{R := \sigma_4.R\{\text{origin} := \sigma_3.R\text{.origin}\}$}}, b, (\lambda \sigma_5, b' \cdot \\
    & \quad \quad \quad \quad \quad \quad \colorbox{yellow!50}{\text{if bdy is precise then}} \\
    & \quad \quad \quad \quad \quad \quad \quad Q(\sigma_5\{h := \sigma_2.h,\colorbox{yellow!50}{$h_? := \sigma_2.h_? \cup \sigma_5.h_?$} \colorbox{orange!30}{$\cup \text{ pred}(\overline{e'})$}\}, b'))))))) \\
    & \quad \quad \quad \quad \quad \quad \colorbox{yellow!50}{\text{else}} \\
    & \quad \quad \quad \quad \quad \quad \quad \colorbox{yellow!50}{$Q(\sigma_5\{h := \sigma_2.h, h_? := \sigma_2.h_? \colorbox{orange!30}{$\cup \text{ pred}(\overline{e'})$}\text{, isImprecise} := \sigma_2.\text{isImprecise} \}, b')))))))$} \\
    & \quad \text{else} \\
    & \quad \quad \text{Let } \text{recunf} \text{ be a fresh function symbol such that} \\
    & \quad \quad \quad 1. \text{ its arity is } |\sigma_1.qvs| \\
    & \quad \quad \quad 2. \text{ it can be applied to the argument vector } \overline{\sigma_1.qvs} \\
    & \quad \quad \quad 3. \text{ its return sort matches } b\text{'s sort} \\
    & \quad \quad Q(\text{recunf} (\overline{\sigma_1.qvs}))
    \end{align*}
}%
\caption{Modified \texttt{eval} rule for \texttt{unfolding} expressions in Gradual Viper}
\label{fig:eval}
\end{figure}
The \texttt{eval} rule for \texttt{unfolding} expressions in Viper \cite{Schwerhoff16} (defined in continuation-passing style) is modified (with highlighted parts) for Gradual Viper. Note, $\sigma$ denotes the symbolic state, $R$ collects run-time checks down a particular execution path, and $Q$ is the continuation (a function that represents the remaining symbolic execution that still needs to be performed).

We choose to retain optimistic heap ($h_?$) chunks differently based on whether \texttt{pred($\overline{e}$)}'s body is precise or imprecise (highlighted in yellow). In the case of a precise predicate body, $h_?$ is not reverted back to the version before the \texttt{consume} call (unlike $h$). Instead, we take the union of $h_?$ before the \texttt{consume} call and $h_?$ after the \texttt{eval $b$} call, thereby regaining chunks lost from the \texttt{consume} call while also retaining chunks assumed during evaluation of $b$. In the case of imprecise predicate body, both $h_?$ and $h$ are reverted back to their respective versions before the \texttt{consume} call.

Additionally, in either case, we add \texttt{pred($\overline{e}$)} to $h_?$ (highlighted in orange), if not already present. This is safe because \texttt{pred($\overline{e}$)} is framed by the \texttt{consume} call within the \texttt{eval} rule for \texttt{unfolding}. 

We add functionality for branch origin tracking (highlighted in blue). Producing \texttt{pred($\overline{e}$)}'s body can potentially cause the execution to branch if the body contains a conditional assertion of the form $e$ ? $\phi_1:\phi_2$ (where $e$ is a boolean expression and $\phi_1$, $\phi_2$ are assertions). We set the origin for the branch in this case to be the \texttt{unfolding} expression. However, if the origin is already set (not \texttt{None}), that means our \texttt{unfolding} expression resides in a method's signature or in a predicate body. In this case, we want the origin to be
the method call or the \texttt{fold}/\texttt{unfold}/\texttt{unfolding} of the aforementioned predicate. So, the origin is not updated.

Note that we omit a \texttt{join} call (present in Viper's \texttt{unfolding eval} rule) since \texttt{join} functionality (joining evaluation branches) is currently not supported in Gradual Viper. 

\end{document}